\documentclass[a4paper,12pt,twoside,titlepage]{article}

\RequirePackage[OT1]{fontenc}
\RequirePackage{amsthm,amsmath}
\RequirePackage{natbib}
\RequirePackage[colorlinks,citecolor=blue,urlcolor=blue]{hyperref}
\usepackage{graphicx}	
\usepackage{amsmath}	
\usepackage{amssymb}
\usepackage[title]{appendix}
\usepackage{pdflscape}	
\usepackage{wrapfig}
\usepackage{lscape}
\usepackage{float}
\usepackage{rotating}
\usepackage[utf8]{inputenc}
\usepackage{booktabs}	
\usepackage{caption}
\usepackage{xcolor}
\usepackage{ulem}
\usepackage{placeins}
\usepackage[ruled,norelsize,vlined]{algorithm2e}
\usepackage[noend]{algpseudocode}      

\usepackage{fancyhdr}
\title{{\LARGE\sc MARGIN-FREE CLASSIFICATION} \\[1ex] {\LARGE\sc AND NEW CLASS DETECTION} \\[1ex] {\LARGE\sc USING FINITE DIRICHLET MIXTURES}\vspace{2cm}}

\author{{\LARGE P. John$^*$, A.R. Brazzale$^*$ and M. S\"uveges$^\dag$}
\\[5ex]
{\Large $^*$University of Padova} \\
{\Large $^\dag$\'Ecole Polytechnique F\'ed\'erale de Lausanne} \vspace{2cm}}

\date{\today\vspace{3cm}}

\begin{document}

\maketitle

\begin{abstract}
We present a margin-free finite mixture model which allows us to simultaneously classify objects into known classes and to identify possible new object types using a set of continuous attributes.  This application is motivated by the needs of identifying and possibly detecting new types of a particular kind of stars known as variable stars.  We first suitably transform the physical attributes of the stars onto the simplex to achieve scale invariance while maintaining their dependence structure.  This allows us to compare data collected by different sky surveys which can have different scales.  The model hence combines a mixture of Dirichlet mixtures to represent the known classes with the semi-supervised classification strategy of \cite{vatanen2012semi} for outlier detection.  In line with previous work on semiparametric model-based clustering, the single Dirichlet distributions can be seen as providing the baseline pattern of the data.  These are then combined to effectively model the complex distributions of the attributes for the different classes.  The model is estimated using a hierarchical two-step procedure which combines a suitably adapted version of the Expectation-Maximization (EM) algorithm with Bayes' rule.  We validate our model on a reliable sample of periodic variable stars available in the literature \citep{dubath2011random} achieving an overall classification accuracy of 71.95\% and a sensitivity of 86.11\% and a specificity of 99.79\% for new class detection.  
\end{abstract}


\section{Introduction}\label{intro}
In many areas of application where statistics is wide\-ly used, such as in digital marketing, social network analysis, functional genomics and proteomics, public health survey and astrophysics, a common goal of the analysis is to assign new observations to predefined classes of similar objects.  At the same time, we may want to pursue the possibility of identifying as yet undiscovered groups in our study population.  Since today's applications often involve large and complex datasets, there is a huge demand of highly efficient techniques for classification and new class detection in terms of flexibility, performance and speed.
In this paper we present a margin-free model for the identification of known and the detection of possibly new classes in a set of unclassified objects motivated by an impellent data analysis problem of modern astronomy. 
This model was developed by considering the most recent supervised and semi-supervised techniques of semi-parametric model-based clustering.

Traditional parametric model-based clustering groups unlabelled data points into different classes using finite mixtures of parametric densities where each component density corresponds to a cluster \citep{fraleyandraftery1998}.  The use of finite mixtures in the context of clustering dates back to the early sixties and has since then seen a wealth of applications \citep{mclachlan2004finite,fruhwirth-schnatter2006,fruhwirth-schnatter2018}.  For a comprehensive review on parametric model-based clustering, see \cite{grun2018} and the references therein.  The most popular choice is to use Gaussian densities as mixture components \citep{mclust}, though finite mixtures of non normal densities have been put forward  with the intent to accommodate for clusters with non Gaussian shapes \citep{lee2013nonGaussian} or to work in non Euclidean spaces such as the sphere \citep{costantin2020sphere}.  Nonetheless, making the decision of which parametric distribution best represents the shapes of the clusters can be difficult especially when we analyze multivariate data.  Furthermore, there is no guarantee that the cluster distributions are unimodal.  

Recent advances in model-based clustering have focused on studying mixture models which combine more flexible cluster densities.  Semi-parametric model-based clustering uses so-called mixtures of mixtures where the class distributions are themselves modelled as a mixture of suitable parametric densities.  The two-level hierarchical structure which characterizes these models is particularly appealing in the clustering and classification context.  The finite mixture models used at the lower level accommodate in a semi-parametric way for possible non Gaussian cluster distributions. They are then combined at the upper level to model the heterogeneous population.  This powerful and very flexible approach has been employed in various ways \citep{bartolucci2005, li2005, di2007mixture, orbanz2005sar}.  Because of the ability of Gaussian mixtures to accurately approximate a wide class of probability distributions, most applications are confined to the case of mixtures of normal mixtures.  Yet, some alternative formulations have started to appear \citep{browne2011model}.

The use of mixture models has also been considered for outlier and anoma\-ly detection which represent key issues in domains as diverse as fraud detection, network security, safety monitoring and recently new physics discoveries.  \cite{eskin2000anomaly} detects computer intrusions using a mixture model which combines a structured distribution for the majority of the data with a uniform distribution and associated small mixture proportion to model the possible anomalies.  Similarly, \cite{lauer2001mixture} explicitly models the existence of outliers in a training dataset using a diffuse Gaussian mixture component.  \cite{ritter1997outliers} develop a heuristic method of parameter estimation in mixture models for data with outliers and design a Bayesian classifier of assignment of $m$ objects to $n\geq m$ classes under constraints. 

A common limitation of these types of models is that they only perform well if the two mechanisms which generated the majority of the data and the anomalies are rather different.  However, the observations which form a possible new cluster need not be outliers by themselves, but their occurrence together is anomalous \citep{Chandolaetal2009anomal}.  This is a typical feature of new class detection.  The recent paper by \cite{vatanen2012semi}, which was motivated by a real data analysis problem of experimental high energy physics, overcomes this difficulty.  A two-step procedure is proposed.  At the first step, labelled background data are classified using a parametric mixture model.  At the second step, unlabelled data are modeled with a mixture of the previously estimated background model and a new class detection model. 


The motivation for this paper is drawn from the most recent advances in observational astronomy.  Current and forthcoming sky surveys, such as Gaia\footnote{https://www.cosmos.esa.int/web/gaia} \citep{GaiaCollaborationPrustietal16}, LSST\footnote{https://www.lsst.org} and Euclid\footnote{https://www.cosmos.esa.int/web/euclid} are designed with the declared intent to collect huge volumes of data.  These consist of 
billions of celestial objects whose nature is unknown and which can belong to fundamentally different type such as stars, quasars, planets, galaxies and diffuse Galactic clouds, just to name a few.  To scientifically exploit the collected databases, we need to label the observed data points.  Our work focuses on the identification and classification of a particular kind of stars known as \textit{variable} stars.  These are stars whose brightness as seen from the Earth changes with time because of various reasons \citep{percy2007understanding}.  Periodic variable stars, in particular, hold a special role in astronomy.  Cepheid variables, for instance, are used to calibrate the cosmological distance scale, while the study of pulsating and binary stars opens insight into the physics of stellar evolution. 

A large body of previous work on the classification of variable stars is available.  \cite{aerts1998discovery} use multivariate discriminant analysis to isolate certain classes of periodic variables.  \cite{debosscher2007automated} implement a procedure for fast light curve analysis, and for the derivation of the attributes which are suitable for variable star classification.  They furthermore propose a classifier based on Gaussian mixtures; see also \cite{debosscher2009automated}.  A Bayesian network classifier is developed in \cite{sarro2009automated}; \cite{blomme2011improved} use multivariate Bayesian statistics and a multistage approach.  \cite{willemsen2007study} present a systematic classification of variable stars using principal components analysis and support vector machines.   \cite{richards2011machine} compare random forest and stochastic gradient boosting for classifying variable stars with noisy time series data.  \cite{dubath2011random} evaluate the performance of a random forest classifier by benchmarking it on the 26 periodic variable star classes listed in the \textit{Hipparcos} catalogue\footnote{https://www.cosmos.esa.int/web/hipparcos/}. 

In addition to the typical difficulties of big data analysis, such as those linked to huge storage volumes and high dimensions, the classification of astronomical objects poses further challenges.  At first, data collected by different sky surveys can have different scales.  This may affect the efficiency of supervised and unsupervised classification models which use data from different sources for training and testing.  We hence must develop models which are independent of the measurement scales used by the surveys; in the remainder of the paper we will refer to these models as \textit{margin-free}. In addition, with so large incoming datasets we cannot rule out the possibility that future surveys will detect new object types.  The \textit{Gaia} mission alone is expected to provide about 18 million variable stars, including 5 million classic periodic variables \citep{eyer2000predictions}.  Both issues need to be tackled.  

The aim of this paper is to build a margin-free model that allows us to simultaneously classify the variable stars of a new dataset into the current known classes and to detect possible new types of variability by using a set of continuous features measured on the stars.  We achieve this by combining a mixture of mixtures model of Dirichlet densities with the semi-supervised classification strategy of \cite{vatanen2012semi}.  The scaling problem is solved by first transforming all attributes used for the classification of the stars to the probability scale, then further to the simplex.  The transformation is carefully chosen so as to maintain the dependence structure among the different features.
%
%
A mixture of Dirichlet mixtures is then fitted to a labeled \textit{background} sample.  
In accordance with previous work on semiparametric model-based classification, the inner mixture models the distribution of the attributes within the classes, while the outer mixture represents the different variable types.  We will call this model the Two-Stage Dirichlet Mixture (TSDM) model.  At last, a \textit{new-class} component is added, which consists of a mixture of additional Dirichlet densities to model the possible anomalies in an unlabelled sample.
This way we not only detect but also describe the patterns of the new variable types.  The global model will be referred to as Fixed-Background model.  
The  model  is  estimated  using  a hierarchical two-step procedure which combines a suitably adapted version of the Expectation-Maximization (EM) algorithm with Bayes’ rule.  We evaluated the performance of the proposed model by validating it on a well-studied subgroup of variable stars of the \textit{Hipparcos} catalogue \citep{perryman1997hipparcos} and compare it with previous work \citep{dubath2011random}.  The model correctly identifies the known variable types with an accuracy of 71.49\% and detects new classes with a sensitivity of 86.11\% and a specificity of 99.79\% yielding an overall accuracy of 71.95\%.

The content of this paper has been arranged as follows. In Section~\ref{modeldefinition}, we present the model, which is estimated according to the algorithm described in Section~\ref{modelfitting}.  Data pre-processing and selection of the most important attributes for classification are discussed in Section~\ref{onlyTSDMHip}.  The results of our analysis are given in Section~\ref{results}.  Section~\ref{conclusion} contains a brief discussion and the conclusions.

\section{Model definition}\label{modeldefinition}
Assume we have a collection of $n$ objects from $K$ known classes and possibly one unknown class that we want to model.  The aim is to predict the class of new instances from the same parent population. Let $\mathbf{Y_1,\ldots,Y_n}$ characterise this random sample of $n$ independent observations, where $\mathbf{Y_i}\in \mathbb{V}^{D-1}$, $i=1,\ldots,n$, takes  values in the open simplex $\mathbb{V}^{D-1}\subset \mathbb{R}^D$ with $D> 1$.  Let $\mathbf{y_1,\ldots,y_n}$ be the observed data points, where $\mathbf{y_i^T}=(y_{i1},y_{i2},\ldots,y_{iD})$ represents the observed value of the \textit{D}-dimensional random vector $\mathbf{Y_i}$ for the $i$th observation.  In our work, the \textit{n}-tuple of points, $\mathbf{y_1,\ldots,y_n}\in\mathbb{R}^D$, represents the $D$ continuous attributes measured on $n$ observations according to which these are to be classified into the $K$ given classes and a possibile new one.  

We solve this problem using the Fixed-Background (FB) model 
\begin{align}\label{eqfb}
f_{FB}(\mathbf{y_i})=(1-\lambda) f_B(\mathbf{y_i})+\lambda f_{NC}(\mathbf{y_i}),
\end{align}
which is defined as a mixture of two components.  The first component, $f_B(\mathbf{y_i})$, represents the so-called \textit{background model} and will be given in Equation~(\ref{tsdmeqn}).  This model is itself a mixture of $K$ densities and represents the marginal distribution of a data point $\mathbf{y_i}$.  The additional component, $f_{NC}(\mathbf{y_i})$, captures deviations from the background model and will be introduced in \S\ref{fbdiscussion}.  We will refer to it as the \textit{new-class model}.  The parameter $\lambda\in(0,1)$ is the new-class mixture probability, that is, the probability that a data point may represent an as yet undetected new class whose density is $f_{NC}(\mathbf{y_i})$.  With model~(\ref{eqfb}) we can both classify incoming new observations according to a pre-specified scheme and detect possible anomalies in the data.  For $\lambda=0$ this reduces to classical model-based classification while for $\lambda=1$ we face a problem in model-based clustering.

As mentioned in the introduction, the shapes of the $K+1$ component distributions may be of a form which is not easily captured by a single parametric model.  Owing to their flexibility, we will use mixtures of $D$-dimensional Dirichlet densities to represent the distributions of the $D$ attributes for both, the $K$ classes of the background model $f_B(\mathbf{y_i})$ and for the new class given by $f_{NC}(\mathbf{y_i})$.  In line with previous work on semiparametric model-based classification, the single Dirichlet distribution can be seen as providing the ``prototype shape'' of the data for varying parameter values.  These are then combined to effectively model the complex distribution of the different classes.  The corresponding mixtures can be of rather different sizes and shapes, depending on the class they are modelling. If each of these were Gamma distributed, their transformation would result in a single Dirichlet distribution of order~$D$.  

Usually, datasets are not characterized by vectors living on the simplex, so we will need to define a mapping from the space of observed variables to the simplex. In \S\ref{transformation2simplex} we will outline how to do this, using a probability integral transform and a special renormalization to preserve the full $D$-dimensional structure.

\subsection{Background model}\label{tsdmdiscussion}
For the $i$th data point $\mathbf{y_i}$ we consider as background model the $K$-component mixture of mixtures model 
\begin{equation}\label{tsdmeqn-outer}
f_B(\mathbf{y_i}; \boldsymbol{\rho}, \boldsymbol{\theta})=\sum_{k=1}^K\rho_k f_k(\mathbf{y_i};\boldsymbol{\theta_k}), \quad \mathbf{y_i}\in \mathbb{V}^{D-1},
\end{equation}
where $\boldsymbol{\rho^T}=(\rho_{1},\ldots,\rho_{K})$ and $\boldsymbol{\theta^T}=(\boldsymbol{\theta_1^T},\ldots,\boldsymbol{\theta_K^T})$ are unknown parameters.  Each component distribution $f_k(\mathbf{y_i}; \boldsymbol{\theta_k})$ is itself a mixture of $J_k$ Dirichlet densities
%
\begin{eqnarray}\nonumber
f_k(\mathbf{y_i};\boldsymbol{\theta_k}) & = &\sum_{j=1}^{J_k}\pi_{kj} f(\mathbf{y_i};\boldsymbol{\alpha_{kj}}) \\
\label{tsdmeqn-inner}
& = & \sum_{j=1}^{J_k}\frac{\pi_{kj}}{\mathbf{B}(\boldsymbol{\alpha_{kj}})}\prod_{d=1}^D y_{id}^{\alpha_{kjd}-1},
\end{eqnarray}
with $\mathbf{y_i}\in \mathbb{V}^{D-1}$.  Here, $f(\mathbf{y_i};\boldsymbol{\alpha_{kj}})$ is the density function of the $j$th Dirichlet distribution in class $k$, which is indexed by the $D$-dimensional parameter $\boldsymbol{\alpha_{kj}}=\{\alpha_{kjd}\}_{d=1:D}$, with $\alpha_{kj} > 0$ for every $k=1,\ldots,K$ and $j=1,\ldots,J_k$.  Furthermore, $\mathbf {B} (\boldsymbol{\alpha_{kj}})=\prod_{d=1}^D\Gamma (\alpha_{kjd})\big/ \Gamma (\sum_{d=1}^D\alpha _{kjd})$, $\Gamma(\cdot)$ being the Gamma function.  Hence, the class-specific parameter vector is $\boldsymbol{\theta_k^T} = (\boldsymbol{\pi_{k}^T}, \boldsymbol{\alpha_{k1}^T}, \ldots, \boldsymbol{\alpha_{kJ_k}^T})$, where $\boldsymbol{\pi_{k}^T}=(\pi_{k1},\ldots,\pi_{kJ_k})$.  To distinguish the component distributions in (\ref{tsdmeqn-outer}) from those in (\ref{tsdmeqn-inner}) we say that the summation over $j=1:J_k$ in Equation~(\ref{tsdmeqn-inner}) represents the \textit{inner mixture} with $J_k$ components, while the summation over $k=1:K$ in Equation~(\ref{tsdmeqn-outer}) is the $K$-component \textit{outer mixture}.  Correspondingly, the parameters $\boldsymbol{\rho}$ and $\boldsymbol{\pi_{k}}$, $k=1,\ldots,K$, are the outer- and inner-mixture probabilities, which by definition satisfy that $\rho_k, \pi_{kj}\in (0,1)$ for all $k$ and $j$, $\sum_{k=1}^K\rho_k=1$ and $\sum_{j=1}^{J_k}\pi_{kj}=1$. 

The resulting background model can be written in the form
\begin{equation}\label{tsdmeqn}
f_B(\mathbf{y_i};\boldsymbol{\rho}, \boldsymbol{\theta})=\sum_{k=1}^K\rho_k\sum_{j=1}^{J_k}\frac{\pi_{kj}}{\mathbf{B}(\boldsymbol{\alpha_{kj}})}\prod_{d=1}^D y_{id}^{\alpha_{kjd}-1}, \quad \mathbf{y_i}\in \mathbb{V}^{D-1}.
\end{equation}
In the following, we will refer to this model as the $K$-component Two-Stage Dirichlet Mixture (TSDM) model.  As the name suggests, the TSDM model imposes a two-level hierarchical structure, which is particularly appealing in the classification context.  Here, $K$ denotes the number of known classes, while $J_k$ represents the number of unknown inner-mixture components which need be used to model the distribution of the attributes for the $k$th class.  Depending on the class it models, $J_k$ can take on rather different values.  That is, each class of the outer mixture is represented by a finite mixture of Dirichlet densities for a total of $K$ outer components.

\subsection{New-class model}\label{fbdiscussion}
For the new class model  
\begin{equation}\label{ncmodel}
f_{NC}(\mathbf{y_i};\boldsymbol{\kappa},\boldsymbol{\beta}) = \sum_{j=1}^{J_{K+1}}\frac{\kappa_{j}}{{\textbf{B}(\boldsymbol{\beta_j})}}\prod_{d=1}^D y_{id}^{\beta_{jd}-1}, \quad \mathbf{y_i}\in \mathbb{V}^{D-1},
\end{equation}
we use again a finite mixture of Dirichlet densities with positive parameters $\boldsymbol{\beta_{j}}=\{\beta_{jd}\}_{d=1:D}$.  As before, $\mathbf {B} (\boldsymbol{\beta_{j}})=\prod_{d=1}^D\Gamma (\beta_{jd})\big/ \Gamma (\sum_{d=1}^D\beta _{jd})$ is the normalizing constant of the $j$th Dirichlet component, $\boldsymbol{\beta^T}=(\boldsymbol{\beta_1^T},\ldots,\boldsymbol{\beta_{J_{K+1}}^T})$ are the Dirichlet parameters and $\kappa_j\in (0,1)$, $j=1,\ldots,J_{K+1}$, are the mixture probabilities.  The value $J_{k+1}$ acts as an upper limit to the number of components in the new-class model.  To make sure that the new-class model $f_{NC}(\mathbf{y_i})$ will well represent the new class, we assume that (\ref{ncmodel}) is an overfitting mixture whose number of components may exceed the number of effectively required prototype distributions.  In our applications we will set $J_{K+1}$ according to scientific input.  

Combining the model definitions given in Equation (\ref{tsdmeqn}) of the TSDM and of the new-class model~(\ref{ncmodel}), the FB model of Equation (\ref{eqfb}) becomes
\begin{align}\nonumber
f_{FB}(\mathbf{y_i})& 
= 
(1-\lambda)\sum_{k=1}^K\rho_k\sum_{j=1}^{J_k}\frac{\pi_{kj}}{\mathbf{B}(\boldsymbol{\alpha_{kj}})}\prod_{d=1}^D y_{id}^{\alpha_{kjd}-1} 
+
\lambda\sum_{j=1}^{J_{K+1}}\frac{\kappa_j}{{\textbf{B}(\boldsymbol{\beta_j})}}\prod_{d=1}^D y_{id}^{\beta_{jd}-1} \\ 
& \label{fbequation}
=
{\lambda_0} f_B(\mathbf{y_i};\boldsymbol{\rho},\boldsymbol{\theta})
+
\sum_{j=1}^{J_{K+1}}\frac{\lambda_j}{{\textbf{B}(\boldsymbol{\beta_j})}}\prod_{d=1}^D y_{id}^{\beta_{jd}-1}, \quad \mathbf{y_i}\in \mathbb{V}^{D-1}.
\end{align}
The parameters $\rho_k$, $\pi_{kj}$, and $\boldsymbol{\alpha_{kj}}$ have the same definitions as in \S\ref{tsdmdiscussion}.  Moreover, we set ${\lambda_0}=1-\lambda$, ${\lambda_j}=\lambda\kappa_j$ and $\sum_{j=1}^{J_{K+1}}{\lambda_j}+{\lambda_0}=1$.

\section{Model fitting}\label{modelfitting}
We estimate the FB model~(\ref{fbequation}) using a hierarchical two-step procedure.  This procedure reflects the peculiar nature of our model, where the background component $f_B(\cdot)$ represents the status quo, that is, the presence of $K$ different and known classes within the current population, while the new-class component $f_{NC}(\cdot)$ accommodates for the possibility that one, or more, new classes may be detected thanks to future observations.  In the first step, which we call the TSDM step, the background model $f_B(\cdot)$ of Equation~(\ref{fbequation}) is trained on suitable training data.  That is, we first estimate the parameters $\boldsymbol{\rho}$ and $\boldsymbol{\theta}$ of the TSDM model~(\ref{tsdmeqn}) using a set of $n_0$ labelled data points $\mathbf{y_i^0}$, $i=1,\ldots,n_0$, for which we know to which class, among the $K$ given ones, they belong to.  In the second step, which we call the FB step, the unlabelled data, which form our dataset of interest, are classified with the Fixed-Background model $f_{FB}(\cdot)$ of Equation~(\ref{fbequation}).  That is, the parameters $\lambda_0$, $\lambda_j$ and $\boldsymbol{\beta_j}$, $j=1,\ldots, J_{K+1}$, of the Fixed-Background model are fitted to the $n$ observations $\mathbf{y_i}$, $i=1,\ldots,n$, however under the constraint of keeping the parameters $\boldsymbol{\rho}$ and $\boldsymbol{\theta}$ of the background model fixed to the estimates obtained at the previous step.  As a by-product, the newly observed data points are either classified into one of the $K$ known classes or used to estimate the new-class distribution $f_{NC}(\mathbf{\cdot})$.  

All parameters are estimated by maximum likelihood with the exception of the vector of outer-mixture probabilities $\boldsymbol{\rho}$ of the TSDM model for which we employ Bayes' rule to update a suitable prior distribution.  A variant of the Expec\-tation-Maximization (EM) algorithm is used to maximize the likelihood functions involved in the TSDM and FB steps with respect to $\boldsymbol{\theta}$, $\boldsymbol{\beta}$ and $\boldsymbol{\kappa}$, which are the Dirichlet parameters of the background and of the new-class model and the mixture probabilities of the latter one, respectively.  The corresponding algorithms are summarized in the two pseudo code Boxes~\ref{code1} and \ref{code2} and will be further discussed in \S\ref{tsdmfitting-1} and \S\ref{fbfitting}.  For ease of notation, the short-hand version $\mathbf{y_i}$ will be used in \S\ref{tsdmfitting} for the training data points $\mathbf{y_i^0}$ and $n$ in place of $n_0$, when estimating the TSDM model.  

\begin{algorithm}[p]
\caption{Estimating the parameters 
of the component distribution $f_k(\mathbf{y_i};\boldsymbol{\theta_k})$ for class $k$ in the TSDM model (Stage~1 of the TSDM step)}
\label{code1}
~
\parbox{0.9\linewidth}{
\begin{itemize}
\item 
\textbf{Step 1 (Model complexity)} : Set the number of components of the inner mixture $f_k(\mathbf{y_i};\boldsymbol{\theta_k})$ to $J_k$. Fix the minimum required number $n_{\rm min}$ of objects in each inner-mixture component.
\item 
\textbf{Step 2 (Initialization)} : Form a vector of initial values for the parameters of the component Dirichlet densities which form the inner mixture and for the inner-mixture probabilities.  
As we are fitting a $J_k$-component mixture, we have $J_k$ sets of $D$-dimensional Dirichlet parameters and $J_k$ inner-mixture probabilities.  These are $\boldsymbol{\alpha^{0}_{k}}=\{\boldsymbol{\alpha^{0}_{kj}}\}_{j=1:J_k}$ and $(\boldsymbol{\pi^{0}_{k}})^T=(\pi^{0}_{k1},\pi^{0}_{k2},\hdots,\pi^{0}_{kJ_k})$ respectively, where $\boldsymbol{\alpha^{0}_{kj}}=\{\alpha^{0}_{kjd}\}_{d=1:D}.$ 
\item 
\textbf{Step 3 (E-step)} : Compute the $Q(\boldsymbol{\theta_k};\boldsymbol{\theta_k^0})$ function defined in Appendix~\ref{app:TSDM}, where $\boldsymbol{\theta_k^T}=(\boldsymbol{\pi_k^T},\boldsymbol{\alpha_k^T})$, using the weights 
$$w_{kj}(\mathbf{y_i};\boldsymbol{\theta^0_k}) = 
\frac{\pi^{0}_{kj}f(\mathbf{y_i};\boldsymbol{\alpha_{kj}^0})}{\sum_{j=1}^{J_k}\pi^{0}_{kj}f(\mathbf{y_i};\boldsymbol{\alpha_{kj}^0})},
$$
for $j=1,\ldots,J_k$.
\item
\textbf{Step 4 (M-step)} : Maximize the $Q(\boldsymbol{\theta_k};\boldsymbol{\theta_k^0})$ function with respect to $\boldsymbol{\theta_k}$ keeping $\boldsymbol{\theta_k^0}$ fixed.  Let $\boldsymbol{\pi_k^1}$ and $\boldsymbol{\alpha_k^1}$ be the corresponding maxima.
\item 
\textbf{Step 5 (Termination step)} : Iterate Steps~3 an 4 by using the maxima $\boldsymbol{\pi_k^1}$ and $\boldsymbol{\alpha_k^1}$ as new initial values.  Terminate the iterative process at step $t$ if 
$Q(\boldsymbol{\theta_k^{t}};\boldsymbol{\theta^{t-1}_{k}}) \leq Q(\boldsymbol{\theta_k^{t-1}};\boldsymbol{\theta^{t-1}_{k}})+\epsilon$, for $\epsilon$ below some preset threshold. 
\item 
\textbf{Step 6 (Checking for local maxima)} : Repeat Steps~2 to 5 for different initial values for fixed $J_{k}$ until convergence. Discard  models that contain less than $n_{\rm min}$ data points. Among the rest, choose the estimates which maximize the likelihood.  
\item 
\textbf{Step 7 (BIC)} : Using the maximum likelihood estimates $\boldsymbol{\pi^{t}_{k}}$ and  $\boldsymbol{\alpha^{t}_{k}}$ identified at Step~6, compute the Bayesian Information Criterion (BIC).
\item 
\textbf{Step 8 (Model selection)} : Repeat Steps~1 to 7 for different values of $J_k$.  Choose the value of $J_k$ which minimizes the BIC.  This yields the final model estimate.
\end{itemize}}
\end{algorithm}

\subsection{Estimating the TSDM model}\label{tsdmfitting}
The estimation of the TSDM model (\ref{tsdmeqn}) poses an identifiability problem as we can permute the $J=\sum_{k=1}^K J_k$ Dirichlet densities $f(\mathbf{y_i};\boldsymbol{\alpha_{kj}})$ in $J!$ different ways, and then group them to form the $K$ different outer-mixture components, without changing the likelihood function.  That is, the TSDM model is not identifiable in the absence of additional information.  One possibility is to impose strong identifiability constraints on the component distributions.  A different strand of literature identifies the $K$ classes after having fitted the mixture model to the data using suitable post-processing procedures.  Identifiability can also be imposed through a suitable prior distribution.  See \cite{malsiner-walli2017postprocessing} and references therein.

We exploit the information included in our training dataset that tells us which observation belongs to which of the $K$ different classes.  The inner- and outer-mixture components of the TSDM model are hence fitted sequentially using a two-stage procedure.  In the first stage we separately fit, for each of the $K$ classes, the finite Dirichlet mixture model (\ref{tsdmeqn-inner}) using maximum likelihood.  The parameters $\boldsymbol{\theta^T_k}=(\boldsymbol{\pi^T_k},\boldsymbol{\alpha_{k1}^T},\ldots,\boldsymbol{\alpha_{kJ_k}^T})$, for $k=1,\ldots,K$, are hence fixed to their estimated values before we move to the second stage.  Here, we form the outer mixture (\ref{tsdmeqn-outer}) by combining the ensemble of all the fitted inner mixtures.  The purpose is to estimate the outer-mixture probabilities $\boldsymbol{\rho}$.  The prior belief of the values $\boldsymbol{\rho}$ can take is expressed through a conjugate prior distribution.  We deliberately switch to the Bayesian paradigm of inference as we want to take account of exogenous information provided by scientific collaborators.  This choice reflects once more the hierarchical structure of the TSDM model where the inner mixtures simply model the complex shape of the $K$ component distributions, while the outer mixture represents the $K$ distinct classes present in the whole population.  

\begin{algorithm}[p]
\caption{Estimating the parameters of the FB model}\label{code2}
~
\parbox{0.9\linewidth}{
\begin{itemize}
\item
\textbf{Step 0 (Background model)} : Set $f_B(\mathbf{y_i}) = f_B(\mathbf{y_i};\boldsymbol{\hat\rho},\boldsymbol{\hat\theta})$ where $\boldsymbol{\hat\rho},\boldsymbol{\hat\theta}$ are the estimates obtained at the TSDM step.
\item  
\textbf{Step 1 (Model complexity)} : Set the number of components of the new-class mixture $f_{NC}(\mathbf{y_i};\boldsymbol{\kappa},\boldsymbol{\beta})$ to $J_{K+1}$.  Fix the minimum required number $n_{\rm min}$ of objects in each inner-mixture component.
\item 
\textbf{Step 2 (Initialization)} : Form a vector of initial values for the parameters of the component Dirichlet densities and for the mixture probabilities.  Since we are fitting a $J_{K+1}$-component mixture, we have $J_{K+1}$ sets of $D$-dimensional Dirichlet parameters plus $(J_{K+1}+1)$ mixture probabilities.  These are $\boldsymbol{\beta^{0}_{j}}=\{\beta^0_{jd}\}_{d=1:D}$ for $j=1,\ldots,J_{K+1}$, $(\boldsymbol{\kappa^0})^T=(\kappa_1^{0},\ldots,\kappa_{J_{K+1}}^{0})$ and $\lambda^{0}$. 
Reparametrize the model into $\boldsymbol{\lambda^T}=\{\lambda_0,\lambda_1,\ldots,\lambda_{J_{K+1}}\}$ where ${\lambda_0}=1-\lambda$ and ${\lambda_j}=\lambda\kappa_j$ for $j=1,\ldots,J_{K+1}$. 
\item 
\textbf{Step 3 (E-step)} : Compute the $Q(\boldsymbol{\psi};\boldsymbol{\psi^0})$ function defined in Appendix~\ref{app:fb}, where $\boldsymbol{\psi^T}=(\boldsymbol{\lambda^T}, \boldsymbol{\beta^T})$ is the vector of all unknown parameters and $\boldsymbol{\psi^0}$ its initial value, using the weights
\begin{small}
$$ w_0(\mathbf{y_i};\boldsymbol{\psi^0}) =
\frac{\lambda_0^0f_B(\mathbf{y_i})}{\lambda_0^0f_B(\mathbf{y_i})+\sum_{j=1}^{J_{K+1}}{\lambda}_j^{0}f(\mathbf{y_i};\boldsymbol{\beta^0_j})} $$
and
$$ w_j(\mathbf{y_i};\boldsymbol{\psi^0}) =
\frac{{\lambda_j}^{0} f(\mathbf{y_i};\boldsymbol{\beta^0_j})}{\lambda_0^0f_B(\mathbf{y_i})+\sum_{j=1}^{J_{K+1}}{\lambda}_j^{0}f(\mathbf{y_i};\boldsymbol{\beta^0_j})} $$
\end{small}
for $j=1,\ldots,J_{K+1}$.
\item 
\textbf{Step 4 (M-step)} : Maximize the $Q(\boldsymbol{\psi};\boldsymbol{\psi^0})$ function with respect to $\boldsymbol{\psi}$ keeping $\boldsymbol{\psi^0}$ fixed.  Let $\boldsymbol{\lambda^1}$ and $\boldsymbol{\beta^{1}}$ be the corresponding maxima.
\item 
\textbf{Step 5 (Termination step)} : Iterate Steps~3 an 4 by using the maxima $\boldsymbol{\lambda^1}$ and $\boldsymbol{\beta^{1}}$ as new initial values.  Terminate the iterative process at step $t$ if $Q(\boldsymbol{\psi^{t}};\boldsymbol{\psi^{t-1}}) \leq 
 Q(\boldsymbol{\psi^{t-1}};\boldsymbol{\psi^{t-1}})+\epsilon$, for $\epsilon$ below some preset threshold.  
\item 
\textbf{Step 6 (Checking for local maxima)} : Repeat Steps~2 to 5 for different initial values for fixed $J_{K+1}$ until convergence.  Discard  models that contain less than $n_{\rm min}$ data points. Among the rest, choose the estimates which maximize the likelihood.  
\item 
\textbf{Step 7 (BIC)} : Using the maximum likelihood estimates $\boldsymbol{\lambda^t}$ and $\boldsymbol{\beta^{t}}$ identified at Step~6, compute the Bayesian Information Criterion (BIC).
\item 
\textbf{Step 8 (Model selection)} : Repeat Steps~1 to 7 for different values of $J_{K+1}$.  Choose the value of $J_{K+1}$ which minimizes the BIC.  This yields the final model estimate.
\end{itemize}}
\end{algorithm}

\subsubsection{Stage~1 of the TSDM step}\label{tsdmfitting-1}
Code Box~\ref{code1} outlines the main steps of the algorithm used to estimate the parameters of the component distribution $f_k(\mathbf{y_i};\boldsymbol{\theta_k})$ for class $k$.  The mathematical details for the E- and M-steps are given in Appendix~\ref{app:TSDM}.  To prevent the EM algorithm to converge to a local optimum, we maximize the likelihood function using different initial values. 
However, no restriction is imposed on the mixture probabilities $\pi_{jk}$. As we do not know a priori how many components need be in the mixture, models of different degree of complexity are estimated using different values of $J_k$.  To avoid overfitting, and overly complex models, we choose the model which minimizes the Bayesian Information Criterion (BIC), as BIC penalizes overparametrized models heavily \citep[\S 2.6]{mclachlanandrathnayake2015,bouveyron-etal2019}. 
The exclusion of degenerate models by requiring that all inner mixtures contain a minimum number of elements (3 for the smallest classes) ensures that the BIC does not decrease below a certain limit and exhibits the common elbow shape.

\subsubsection{Stage~2 of the TSDM step}
In the second stage, the TSDM model is formed according to Expression (\ref{tsdmeqn-outer}) by combining the densities $f_k(\mathbf{y_i};\boldsymbol{\hat\theta_k})$ of the $K$ inner mixtures indexed by the parameters estimated at the first stage.  Here, we use the notation $\boldsymbol{\hat\theta_k}$, $k=1,\ldots,K$, to highlight that the parameters $\boldsymbol{\theta_k}$ are to be kept fixed to the estimates obtained at the previous stage.  To estimate $\boldsymbol{\rho}$ we proceed as follows.   Let $\mathbf{y}=(\mathbf{y_1},\ldots,\mathbf{y_{n}})$ be the training dataset.  Let $\mathbf{s^T}=(s_1,\ldots,s_{n})$ be the corresponding set of variables which specify to which particular class, among the $K$ given ones, the observations $\mathbf{y_i}$ belong to.  That is,
\begin{equation}
\label{indicatorS}
\mathbf {I}(s_i=k) = {\begin{cases}
            1 & {\text{if }}\quad \mathbf{y_i} \in \text{class $k$} \\ 
            0 & {\text{if }}\quad \mathbf{y_i} \notin \text{class $k$}. 
                         \end{cases}} 
\end{equation}                         
%
Note that while in the first stage we used the mixtures for a density estimation within each class, at this second stage we use the values of the indicator variables (\ref{indicatorS}) to define the second-level mixtures and perform a supervised classification.  At the first stage, the assignment of $\mathbf{y_i}$ to a particular Dirichlet distribution within the inner mixture was only a technical artifact used to maximize the likelihood function.
 
The complete-data likelihood $L(\boldsymbol{\rho}, \boldsymbol{\hat\theta}\mid\mathbf{y},\mathbf{s}) = f(\mathbf{y},\mathbf{s} \mid \boldsymbol{\rho}, \boldsymbol{\hat\theta})$ can be written as
\begin{align}
\nonumber
L(\boldsymbol{\rho}, \boldsymbol{\hat\theta}\mid\mathbf{y},\mathbf{s})
  & =\prod_{i=1}^n\prod_{k=1}^K \left\{\rho_k f_k(\mathbf{y_i}\mid\boldsymbol{\hat\theta_k})\right\}^{\mathbf {I}(s_i=k)} \\
\nonumber
  & = \left\{\prod_{k=1}^K \left[\prod_{i:s_i=k}f_k(\mathbf{y_i}\mid\boldsymbol{\hat\theta_k}) \right]\right\} \times \left(\prod_{k=1}^K\rho_k^{n_k}\right) \\
  & \propto \left(\prod_{k=1}^K\rho_k^{n_k}\right),
\label{eq4.2}
\end{align}
where $n_k=\sum_{i=1}^{n} \mathbf {I}(s_i=k)$ for $k=1,\ldots,K$ represents the number of data points which belong to class $k$.  Because of the constraint $\sum_{k=1}^K \rho_k=1$, the group sizes, $n_k$, represent a sample from the multinomial distribution.  Expression~(\ref{eq4.2}), when regarded as a function of $\boldsymbol{\rho^T}=(\rho_1,\rho_2,\hdots,\rho_K)$, is hence proportional to the likelihood of a multinomial random variable with probability vector $\boldsymbol{\rho}$ whose conjugate prior is a Dirichlet distribution.  Let 
$$f(\boldsymbol{\rho}\mid\boldsymbol{e}) = 
\frac{1}{\mathbf{B}(\boldsymbol{e})}\prod_{k=1}^K \rho_{k}^{e_{k}-1}, \quad \boldsymbol{\rho} \in \mathbb{V}^{K-1},$$
be the prior for $\boldsymbol{\rho}$ with Dirichlet parameters $\boldsymbol{e}=({e_{1}},\ldots,{e_{K}})$.  These are in practice set after input from collaborators with expertise in the application domain. The posterior distribution for the outer-mixture probabilities $\boldsymbol{\rho}$,
\begin{align*}
f(\boldsymbol{\rho}\mid\mathbf{y}, \mathbf{s};\boldsymbol{e}) 
\propto \prod_{k=1}^K \rho_k^{n_k+e_k-1},
\end{align*}
corresponds to a Dirichlet distribution with parameters $e^\prime_k=e_k+n_k$ for $k=1,\hdots,K$.  The mode of this posterior distribution is used as a Bayesian point estimate of the outer-mixture probabilities $\boldsymbol{\rho}$.

\subsection{Estimating the FB model}\label{fbfitting}
Similarly to the TSDM model, the FB model is fitted to the unlabelled data $\mathbf{y_i}$, $i=1,\ldots,n$, by maximizing the likelihood of model~(\ref{fbequation}), however under the constraint that the parameters of the background model $f_B(\mathbf{y_i};\boldsymbol{\hat\rho},\boldsymbol{\hat\theta})$ are kept fixed to the values estimated in the previous step.  The pseudo code Box~\ref{code2} outlines the main steps of the EM algorithm, while the mathematical details are given in Appendix~\ref{app:fb}.  Again, we do not want to select a model that is overly complex and hence select the best model by minimizing the BIC.  Moreover, to avoid the EM algorithm to converge to a local optimum, we evaluate the maximum for different initial values of $\boldsymbol{\psi^T}=(\boldsymbol{\lambda^T}, \boldsymbol{\beta^T})$.  Here, $\boldsymbol{\lambda^T}=(\lambda_0,\lambda_1,\ldots,\lambda_{J_{K+1}})$ is as defined at (\ref{fbequation}) and $\boldsymbol{\beta^T}=(\boldsymbol{\beta_1^T},\ldots,\boldsymbol{\beta_{J_{K+1}}^T})$ are the Dirichlet parameters of the new-class model.  No restriction is imposed on the mixing probabilities $\kappa_j$, $j=1,\ldots,J_{K+1}$, and hence on the $\boldsymbol{\lambda}$'s.

\begin{table}[t]
\centering
\caption{The 23 variability types included in our validation dataset listed together with their acronym, the corresponding number of instances and reference.  The dataset contains 1,661 data points in total; selection of these data points was based on the type-assignment process detailed in \cite{dubath2011random}.  The reference column reports the data source which are: the two catalogues \textit{Hipparcos} and AAVSO and personal communications (p.c).}
\label{table_alltypes}
\resizebox{11cm}{!}{
\begin{tabular}{@{}llcl@{}}
\hline
\textbf{Type}                      &  \textbf{Acronym}      & \textbf{Instances}  & \textbf{Reference}     \\ \hline
Eclipsing binaries        & EA     & 228  & Hipparcos          \\
                          & EB     & 255  & Hipparcos          \\
                          & EW     & 107  & Hipparcos          \\
Ellipsoidal               & ELL    & 27   & Hipparcos          \\
Long period               & LPV    & 285  & Lebzelter (p.c)   \\
RV Tauri                  & RV     & 5    & AAVSO              \\
W Virginis                & CWA    & 9    & AAVSO              \\
                          & CWB    & 6    & AAVSO              \\
Delta Cepheid             & DCEP   & 189  & AAVSO              \\
~~~(first overtone)          & DCEPS  & 31   & AAVSO              \\
~~~(multi-mode)              & CEP(B) & 11   & AAVSO              \\
RR Lyrae                  & RRAB   & 72   & AAVSO              \\
                          & RRC    & 20   & AAVSO              \\
Gamma Doradus             & GDOR   & 27   & De Cat (p.c)      \\
Delta Scuti               & DSCT   & 47   & AAVSO              \\
~~~(low amplitude)           & DSCTC  & 81   & AAVSO              \\
Beta Cephei               & BCEP   & 30   & De Cat (p.c)      \\
Slowly Pulsating B star   & SPB    & 81   & De Cat (p.c)      \\
B emission-line star  and \\ 
Gamma Cassiopeiae    & BE+GCAS    & 13    & AAVSO              \\
Alpha Cygni               & ACYG   & 18   & AAVSO              \\
Alpha-2 Canum Venaticorum & ACV    & 77   & Romanyuk (p.c)   \\
SX Arietis                & SXARI  & 7    & Romanyuk (p.c)   \\
RS Canum Venaticorum and \\
BY Draconis               & RS+BY     & 35    & \cite{eker2008vizier} \\ \hline
TOTAL                          & & 1,661 &                   
\end{tabular}
}
\end{table}

Once the FB model is fitted, all observations can be assigned to one of its components by maximum a posteriori classification \citep{mcnicholas2011}.  We will use this feature in \S\ref{results} to quantify the performance of our model. In practice, we first determine whether a data point $\mathbf{y_i}$ belongs to the background model $f_B(\cdot) = f_B(\cdot;\boldsymbol{\hat\rho},\boldsymbol{\hat\theta})$ or to the new-class model $f_{NC}(\cdot)$ according to whether 
$$
\mathrm{Pr}\left(S_i=0 \mid \mathbf{Y_i}=\mathbf{y_i}; \boldsymbol{\hat\lambda},\boldsymbol{\hat\beta}\right) =
\frac{\hat\lambda_0 f_B(\mathbf{y_i})}{\hat\lambda_0 f_B(\mathbf{y_i})+\sum_{j=1}^{J_{K+1}} \hat\lambda_j f(\mathbf{y_i};\boldsymbol{\hat\beta_j})} 
$$
is maximum.  With a slight abuse of notation, here the variable $S_i$ assumes the value 0 if the observation belongs to the Fixed-Background model and 1 if it belongs to the new-class model.  If the data point $\mathbf{y_i}$ gets assigned to the fixed background, we can further determine the class it belongs to by assessing which $k$ maximizes the probability
\begin{equation}
\label{Bayesclass}
\mathrm{Pr}\left(S_i=k \mid \mathbf{Y_i}=\mathbf{y_i}; \boldsymbol{\hat\rho},\boldsymbol{\hat\theta}\right) =
\frac{\hat\rho_k f_k(\mathbf{y_i};\boldsymbol{\hat\theta_k})}{\sum_{k=1}^K \hat\rho_k f_k(\mathbf{y_i};\boldsymbol{\hat\theta_k})}.
\end{equation}  
Here, $S_i$ is the variable defined on page~\pageref{indicatorS} which pinpoints the class, among the $K$ given ones, from which the data point $\mathbf{y_i}$ arose.  This so-called Bayesian classifier minimizes the expected misclassification rate \citep{mclust}.      


\section{Application to \textit{Hipparcos} data}\label{appl2hippdata}
Our research is motivated by an impellent data analysis problem of modern astronomy.  Several of the ongoing sky surveys are in the process of observing and collecting huge amounts of data on millions if not billions of target objects.  We are interested in a particular type of objects called variable stars.  These are stars whose brightness as seen from the Earth changes with time because of various reasons: the luminosity of the star may actually fluctuate, or the changes in brightness may be only apparent and caused by an object that partly blocks the amount of light which reaches Earth. 
Periodic variables, in particular, play a central role in our understanding the formation of the Universe.  Many of the current surveys are in the final phase of their timeline.  The development of automated algorithms for the classification of these types of stars according to their most relevant features and for the discovery of possible new star classes is becoming of primary importance.

\subsection{Dataset used}\label{datasetused}
We validated model~\eqref{fbequation} on the sample of periodic variable stars considered in \cite{dubath2011random}, which give the first systematic, fully automated classification of this kind of celestial objects.  Our dataset consists of 1,661 stars which belong to 23 different types as listed in Table~\ref{table_alltypes} together with their acronym, the number of instances for each of the individual types and their reference.  These data were mainly sourced from the \textit{Hipparcos} catalogue \citep{perryman1997hipparcos}, which lists nearly 120,000 stars observed with the highest precision by ESA's \textit{Hipparcos} satellite launched in 1989.  Of these we use a subset of 1,661 strictly periodic variables.  
The data were further revised with recent information from the International Variable Star Index catalogue \citep{watson2009vizier} issued by the American Association of Variable Star Observers (AAVSO). In addition, some types with too few instances, or large similarities to each other, were combined together to form a single class.  Types from personal communications were also included in the dataset, as it was based on experience by specialists.  For more details on the dataset formation, we refer the interested reader to \cite{dubath2011random}.

\begin{sidewaysfigure}[p]
\vspace{-12pt}
\centering
\begin{tabular}{c} 
     \includegraphics[width=0.5\textwidth]{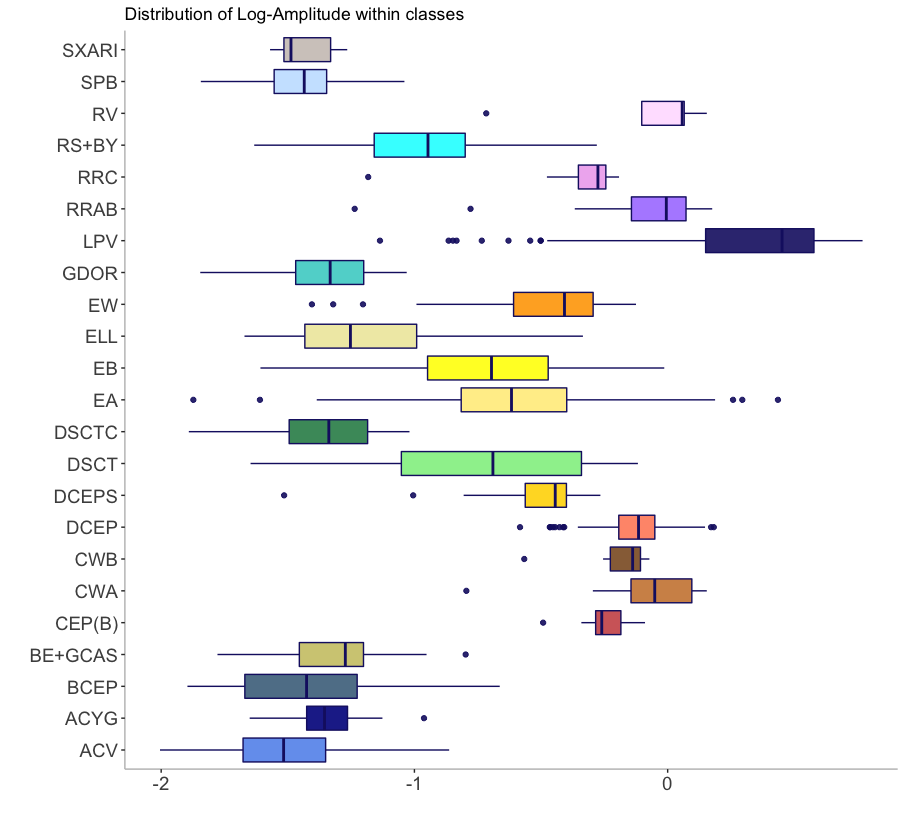}
     \includegraphics[width=0.5\textwidth]{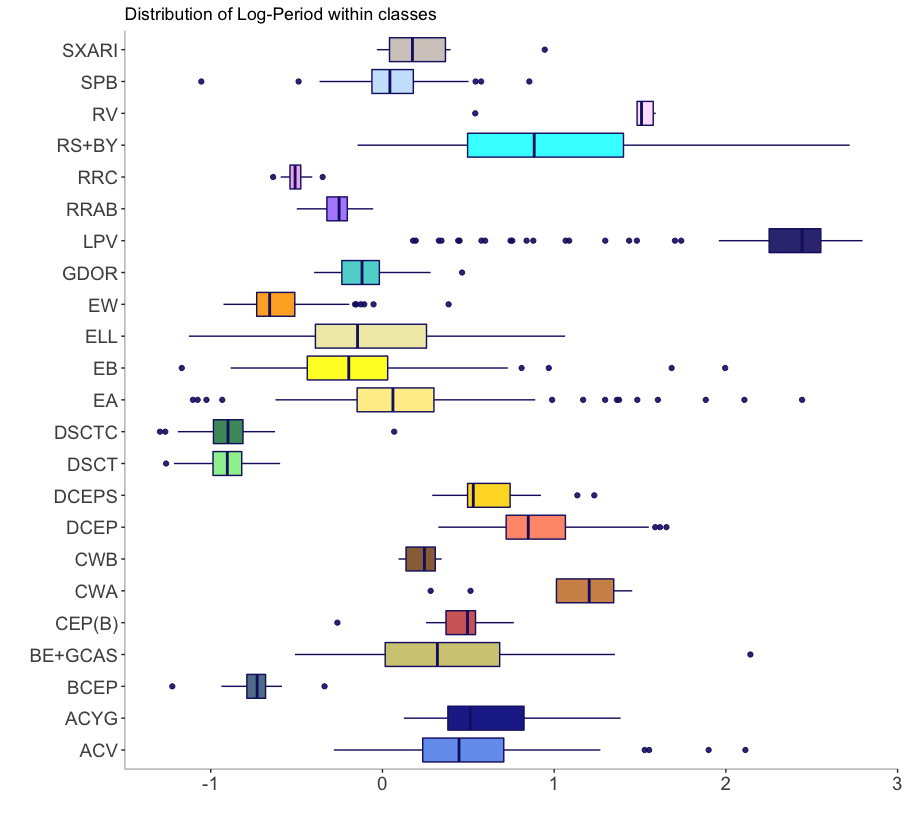}
\end{tabular}
\caption[]{\footnotesize Boxplots of the two attributes amplitude (left) and period (right) on the base 10 logarithmic scale for the 1,661 stars split up into the 23 variable classes considered in this work.  The complete list of all attributes is given in Table~\ref{table_allattributes}.
} 
\label{logperlogamp_boxplot}
\end{sidewaysfigure}

\begin{figure}[t]
\vspace{-12pt}
\centering
\begin{tabular}{c} 
     \includegraphics[width=1\textwidth]{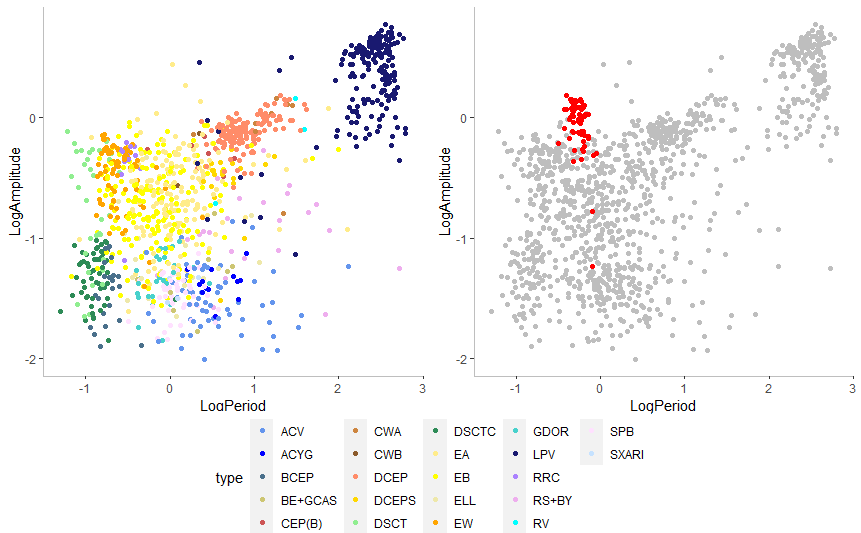} 
\end{tabular}
\caption[]{\footnotesize The decadic LogPeriod--LogAmplitude scatterplots for the training data set (left) and with the putative new class RR Lyrae (RRAB) added as red points (right).  
The complete list of all attributes is given in Table~\ref{table_allattributes}.  Details on the variability class are given in Table~\ref{table_alltypes}. 
} 
\label{logperlogamp_scatter}
\end{figure}

Variable stars are typically classified according to a number of features which include photometric attributes of the emitted light curves.  In astronomy, a light curve charts the light intensity of a celestial object as a function of time.  For variable stars, the shape of the light curve gives valuable information about the global stellar characteristics which determine the fluctuations in brightness.  
Our dataset considers the 42 photometric attributes listed in Table~\ref{table_allattributes}.  For the sake of illustration, Figure~\ref{logperlogamp_boxplot} summarizes the distributions of the two attributes which determine the amplitude (LogAmplitude) and the period (LogPeriod) of the light curve on the base 10 logarithmic scale for the 1,661 stars split up into the 23 variable classes listed in Table~\ref{table_alltypes}.  As our goal is to investigate the ability of our model to both, correctly classify the variables stars into their known classes and to detect a possible new type, the 72 RR Lyrae (RRAB) variables were treated as forming a fictive ``new'' class.  
Figure~\ref{logperlogamp_scatter} gives the decadic LogPeriod--LogAmplitude scatterplots for the 22 ``known'' variable classes which form the background model (left) and with the putative ``new'' class RRAB added as red points (right).

\subsection{TSDM and FB model fits}\label{onlyTSDMHip}
\begin{figure}[h]
\centering
\includegraphics[width=1\textwidth]{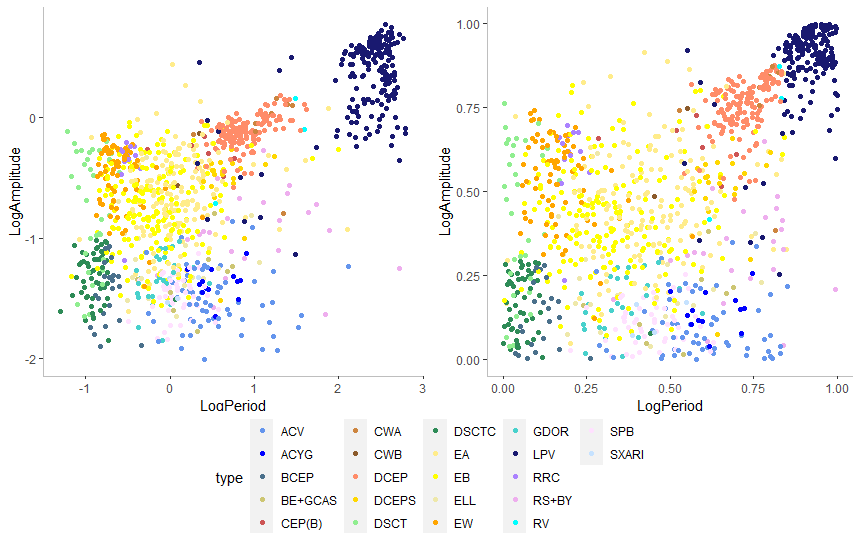}
\caption[]{\footnotesize The decadic LogPeriod--LogAmplitude scatterplots for the raw data which form the training dataset (left) and after transforming them to the simplex (left).
}
\label{STT1_STT2}
\end{figure}

A key requirement of the margin-free classification and new-class detection model developed in Section~\ref{modeldefinition} is that it maintains the dependence structure of the original data.  In the astronomical context, this means that a classifier built using model~\eqref{fbequation} for a specific set of photometric filters can be used to predict the classes of data points observed with a different set of filters as long as the dependence structure of the new data is sufficiently close to the training data.  Furthermore, there may be several alternative ways to characterize a given physical property of the star, which is why many of the original attributes, such as the JmK and JmH or the V-I and B-V color indices, are typically highly correlated (Pearson correlation $> 0.8$).  To lower CPU requirements and increase accuracy, we restricted our analysis to the 16 most meaningful attributes which are highlighted in boldface in Table~\ref{table_allattributes} as done by \cite{dubath2011random}.  We then suitably transformed the original continuous features onto the simplex.  The corresponding details are given in Appendices~\ref{selectionofattributes} and \ref{transformation2simplex}.  Figure~\ref{STT1_STT2} shows the decadic LogPeriod--LogAmplitude scatterplots of the raw measurements for the ``known'' 22 variable classes (left) and after transforming them to the simplex (right).  As we can see, the dependence structure is maintained but the marginal distributions are now mapped to the $(0,1)$ support.

Model~\eqref{fbequation} was hence trained on the transformations of the 16 selected attributes using a random subsample which includes 70\% of the instances available for the 22 variable classes which form the background model.  The 30\% of held-out data were used to evaluate the performance of the fitted Fixed-Background model.  As outlined in Section~\ref{modelfitting}, in the first stage of the TSDM step, a Dirichlet mixture was fitted to each class; see code Box~\ref{code1}.  This yielded the estimates of the inner-mixture parameters  $\boldsymbol{\theta^T_k}=(\boldsymbol{\pi^T_k},\boldsymbol{\alpha_{k1}^T},\ldots,\boldsymbol{\alpha_{kJ_k}^T})$, $k=1,\ldots,22$.  The number of components $J_k$ varies between a minimum of 1, for the very small classes, and a maximum of 7.  In the second stage, Bayes's rule gave the outer-mixture weights $\rho_1,\ldots,\rho_{22}$ of the background model using a non-informative Dirichlet prior of the form $\rm{Dir}(1/22, 1/22, \hdots,1/22)$.  The latter model, with its parameters fixed to the estimates obtained in Step~1, served in the second step as the background component of the FB model to estimate the parameters $\boldsymbol{\theta^T_{23}}=(\boldsymbol{\kappa^T},\boldsymbol{\beta_1^T},\ldots,\boldsymbol{\beta_{J_{23}}^T})$ of the new-class model and the new-class weight $\lambda$; see code Box~\ref{code2}.  In all, 1,053 parameters were estimated.

\begin{figure}[p]
\centering
\vspace{-10pt}
\includegraphics[width=0.9\textwidth]{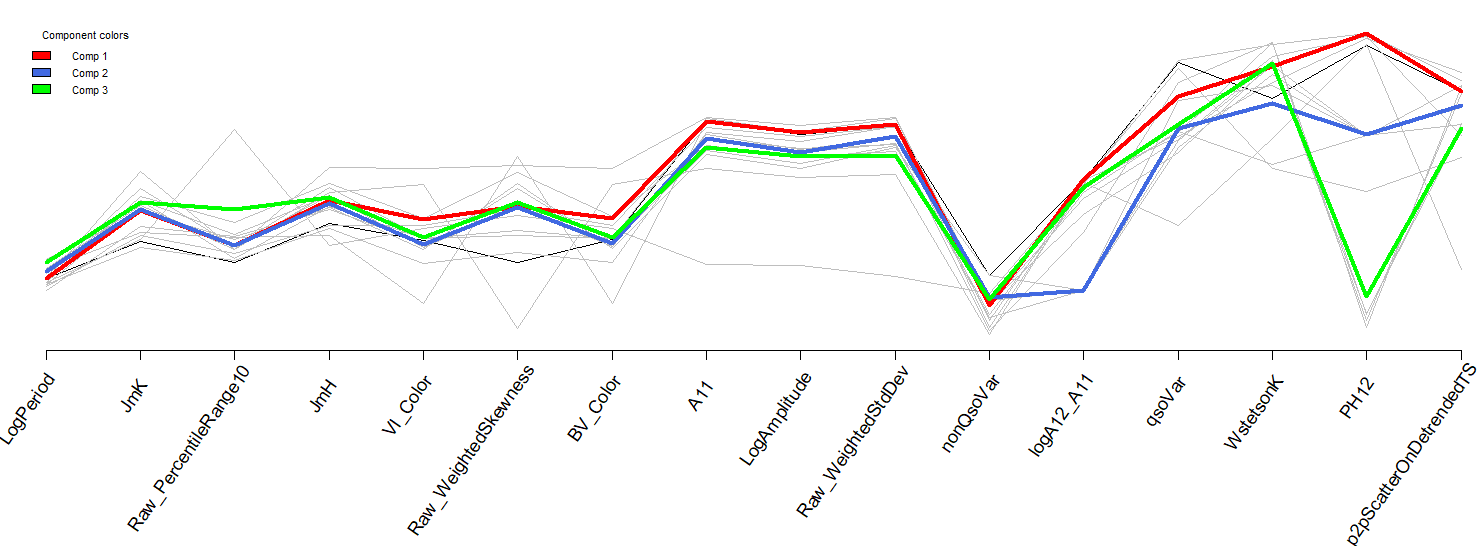}\vspace{-2pt}
\includegraphics[width=0.9\textwidth]{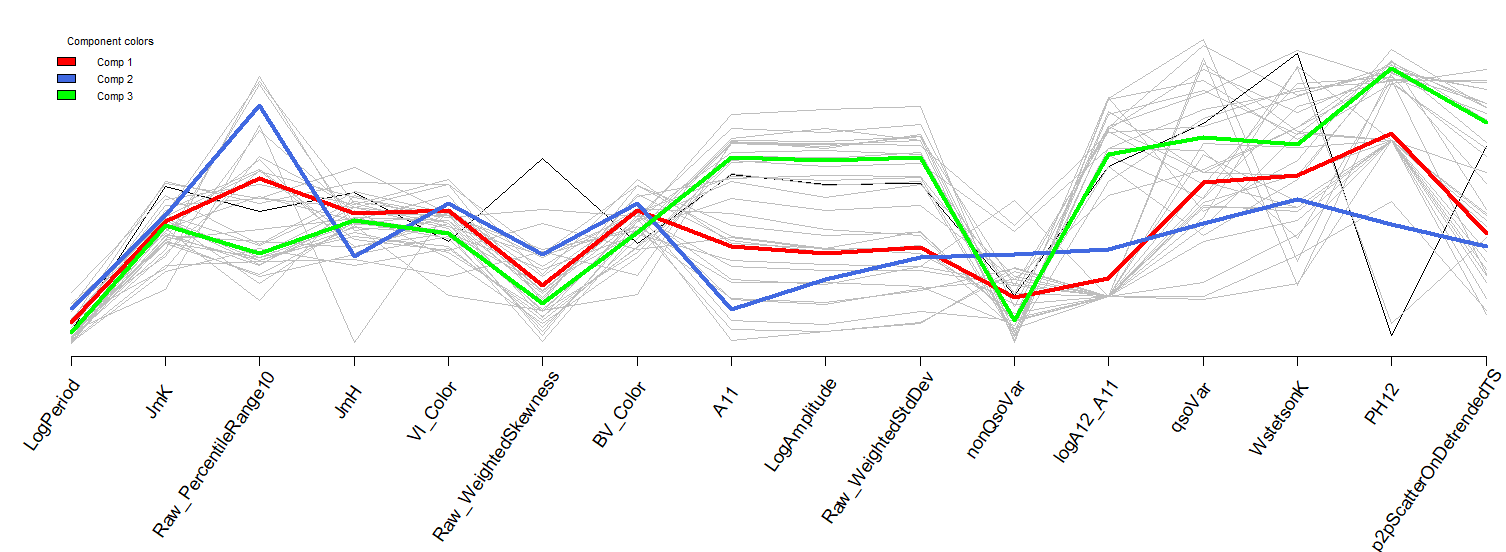}\vspace{-2pt}
\includegraphics[width=0.9\textwidth]{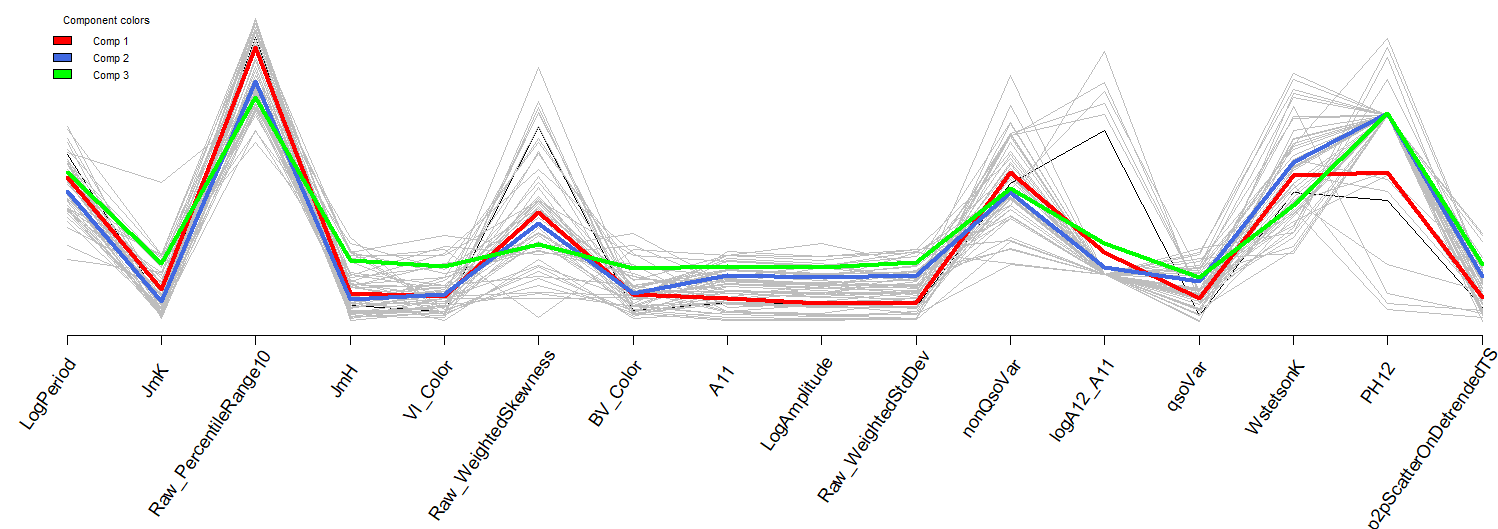} \vspace{-2pt}
\includegraphics[width=0.9\textwidth]{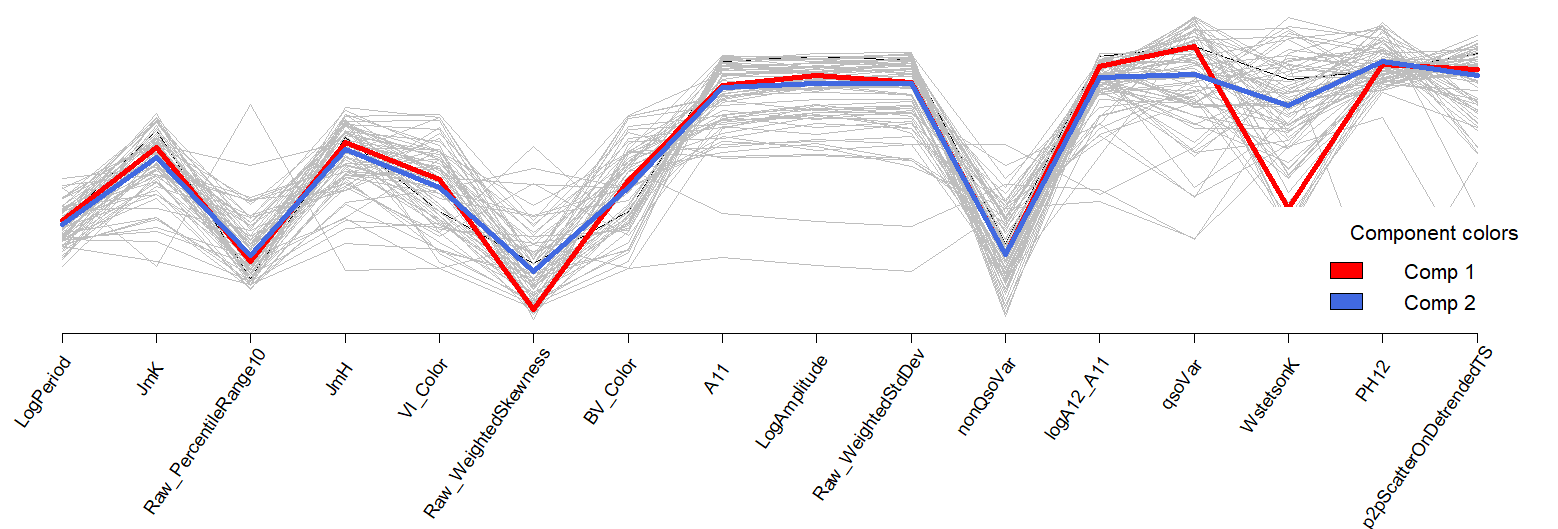}
\vspace{20pt}
\caption[]{{\footnotesize The signatures of each estimated component against the training data (in gray) for RR Lyrae (RRC), Delta Scuti (DSCT) and Slowly Pulsating B (SPB) variables (first three, from top to bottom).  The signatures of the ``newly'' detected class, RRAB, against the testing data is shwon at the bottom.  RRAB is detected at an accuracy of 86.11\%. \\}}
\label{tsdm_stt2_lpv_signatures}
\end{figure}


We can visually evaluate the performance of our FB model fit by inspecting the \textit{signatures} of the Dirichlet components which form the 22 inner mixtures of the TSDM model. 
These are defined as the broken-line plots of the means
$$\left\{\hat{\alpha}_{kjd} / \sum_{d=1}^D \hat{\alpha}_{kjd}\right\}_{d=1:D}, \quad j=1,\ldots,J_k, \ \ k=1,\ldots,22, $$ 
of the Dirichlet components, where $\boldsymbol{\hat{\alpha}_{kj}^T}=(\alpha_{kj1},\ldots,\alpha_{kjD})$ are the parameter estimates of the $j$th  Dirichlet density in the $k$th inner mixture.  The same can be done for the new-class model.  The top three panels of Figure~\ref{tsdm_stt2_lpv_signatures} show the signatures (in red, blue and green) of the three Dirichlet components against the training data (in gray) for the small variable class Delta Scuti, the medium-sized RR Lyrae class and for the large class of Slowly Pulsating B stars.  The signatures of the two components (in red and blue) which form the ``newly'' detected class, RRAB, against the testing data (in gray) is shown at the bottom. The close agreement between the fitted model parameters and the observed data points suggests a good performance of the model.  

In the next section, we benchmark our Fixed-Background model with the classification results of \cite{dubath2011random} who use a random forest classifier on 14 attributes which largely overlap with ours.


\section{Results}\label{results}
As discussed in Section~\ref{modeldefinition}, 
we expect the ``known'' classes to be classified by the background model, and the ``unknown'' RRAB class to be detected by the new-class model.  Two comparisons will be made, one for the TSDM model trained on the 22 given classes, and a second to evaluate the performance of our margin-free model to detect the RRAB class.

\subsection{Assessment of the TSDM model}\label{resultsTSDM}
\begin{figure}[p]
\centering
\includegraphics[width=0.68\textwidth]{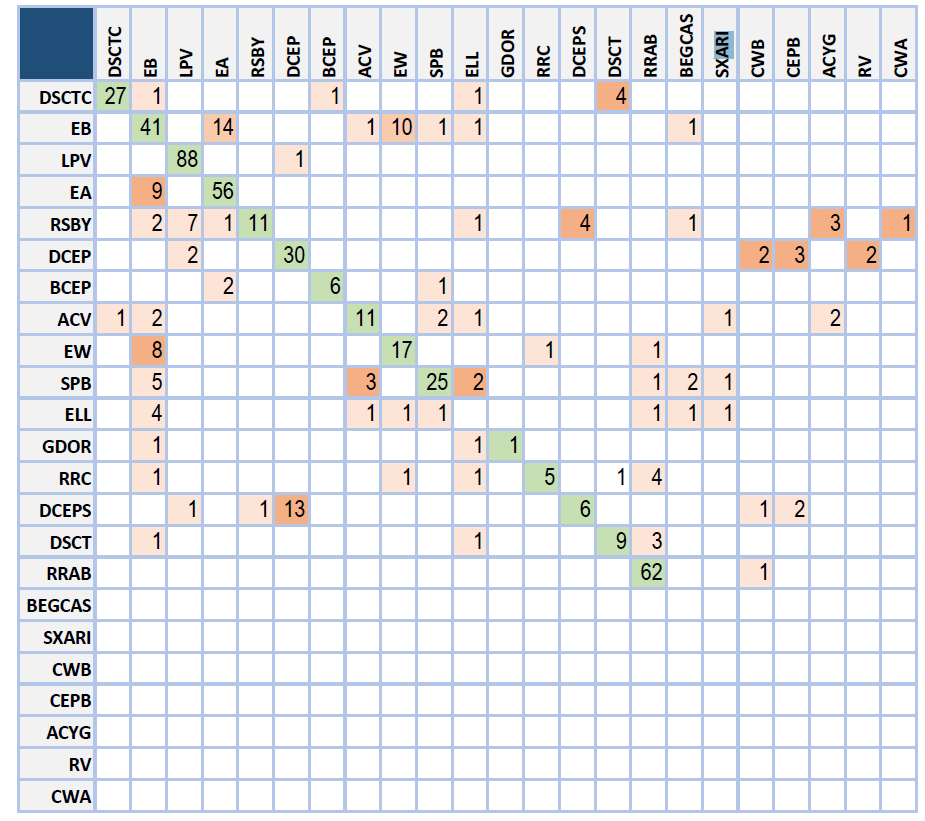}\\ \includegraphics[width=0.70\textwidth]{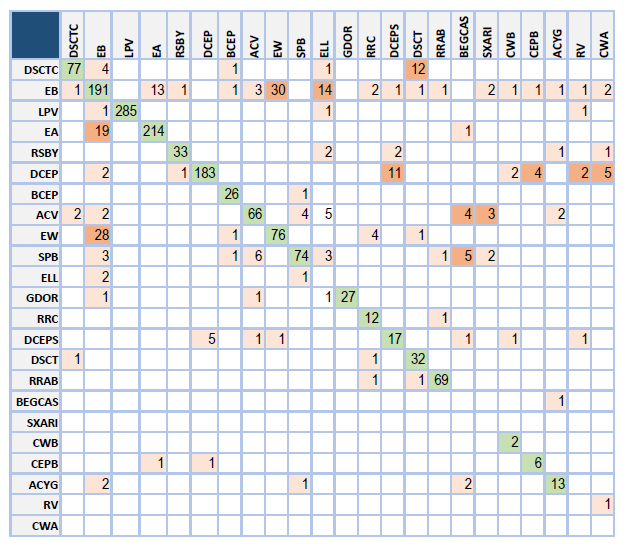}
\caption[]{\footnotesize The confusion matrix evaluated on the test set for our Fixed-Background model (top) and the confusion matrix of \cite{dubath2011random} (bottom).   
}
\vspace{-16pt}
\label{confusions}
\end{figure}

The confusion matrix in Figure~\ref{confusions} shows how the data points of the 22 classes and of the RRAB class are classified by the background model and by the new-class model, respectively.  Leaving out the RRAB and comparing only our background model to the same subset of the confusion matrix in \citet{dubath2011random}, we find an overall accuracy of 67.7\% for our background model, and 84.3\% for the random forest model.  The difference is likely to be caused by a number of effects.  First, \citet{dubath2011random} used the whole dataset as the training set, since with random forest assessment of the model performance is done on-the-fly using out-of-bag samples \citep{breiman2001random}.  
The division of the data into a training and a test set implies that our model is trained (and also tested) on fewer objects, which makes the trained model less precise.  Second, the TSDM model needs to be fitted separately to each class, some of which have as few as 5 objects, whereas random forest uses all data points simultaneously to build a joint model for all classes.  The Dirichlet mixtures therefore can be expected to have worse performance on small classes than a random forest clasifier, which is what we see in Figure~\ref{confusions}, for instance, for the classes SXARI, CWA, CWB, ACYG or CEPB.

However, the pattern of misclassification is  similar in the TSDM and the random forest models, and have reasons originating in astrophysics, suggesting that the model provides a physically reasonable classification. In both models, most confusion arises among the classes within three supergroups: (i) multiple star systems (EA, EB, EW and ELL), (ii) Cepheid-type pulsating variables (DCEP, DCEPS, CEPB, CWA, CWB and RV), and (iii) bright blue variables (ACYG, SPB, BEGCAS and ACV). The star types are confounded in majority with other classes only in their supergroup, and there is only a small confusion between other classes and these supergroups. The main cause for this confusion is that these groups contains subtypes of a single variable family with similar physics behind their variability, or their members are relatively similar to each other in terms of their light curve parameters and astrophysical attributes.
Take for instance the case of eclipsing binaries and ellipsoidal variables (EA, EB, EW, ELL). The class-wise accuracies of the FB model for these are 77\%, 55\%, 59\% and 0\%, respectively. About 19\% of the EA’s were misclassified to EB’s, and 34\% of the EW’s to EB. None of the ELL variables was classified correctly. These fractions are similar to those found by \citet{dubath2011random}. 

In conclusion, although our mixture model performs somewhat less well in the supervised classification part than random forest classifier, its misclassification is reasonable in the sense that it is the consequences of real astrophysical or observational similarities in the variable stars. Thus, the Dirichlet mixture classifier reflects well the astrophysical reality of the variable star population, and can be used as a background model against which new, interesting classes could be detected.

\subsection{Assessment of the FB model} \label{resultsFB}

The FB model detects the RRAB class at an accuracy of 86.11\%, which is an excellent result. When three different training-test set combinations of the data were used, the detection accuracy showed a variation of up to 4\% . The classification specificity of the model is 99.79\%. 
Out of the 72 RRAB data points, 10 were misclassified as DSCT, RRC, SPB, EW and ELL.  
The overall classification accuracy for the complete set of 23 classes is 71.95\%, an improvement on the accuracy of the background model containing 22 classes, due to the fact that the RRAB class is comparatively easy to distinguish, and the class is well represented with its 72 objects.

To evaluate the stability of the shown results, we investigated some alternative formulations of the background model in combination with different selections of the attributes.  As mentioned in Section~\ref{onlyTSDMHip}, some of the attributes have correlations $> 0.8$.  We also explored the performance of our classification model using a subset of attributes which are much less correlated.
The classification accuracy increased by a meagre 0.60, to 72.09\%, largely owing to the better classifications in the eclipsing binaries EA and EB. This corroborates our decision to use the 16 attributes of Section~\ref{selectionofattributes}.

\section{Discussion}\label{conclusion}
Motivated by the advent of large surveys in astronomy and the huge influx of data containing potentially new, rare physical phenomena, we formulated a semi-supervised classification model aimed at the detection of novelties in these data.
Our model consists of two steps: first, we model the known background (our TSDM model), and second, we join this known background model to a flexibly specified mixture model for the novelties, in order to estimate the parameters of the latter and detect the novel phenomena in the data set (the FB model).

To construct the mixtures, we used Dirichlet distributions as the fundamental building blocks instead of the most commonly used Gaussian or $t$-distributions. This is motivated by the fact that in astronomy, there is a wide variety of instruments gathering data from many similar collection of objects. Such closely related, but not identical mappings of the objects into numerical data means that we are often dealing with data sets that have the same dependence structure (the same {\it copula}, speaking in statistical terminology), but different margins. In this case, it is advantageous to build classification models directly on the dependence structure, and not to rely on the margins. To achieve this, we applied the probability integral transformation to the margins, in combination with a nonparametric smoother, and normalised the resulting vector to map onto the unit simplex.

Despite the motivation for using margin-free classification, we also explored the possibility of using multivariate Gaussian densities in a similarly constructed Two-Stage Gaussian Mixture (TSGM) model. We used the R package {\ttfamily mclust} \citep{mclust} to fit the data 
which gave a lower accuracy of 59.52\%.  This model also tended to classify the larger classes, with more than 50 data points in the training dataset, like LPV, DCEP, DSCTC, EA and EB, well. However, 13 out of 23 classes including DSCT, ACV, GDOR, RRC and BCEP were not detected at all. With low detection ability of the smaller classes, the TSGM model would exhibit low classification accuracy for new class detection.

Our combined Fixed-Background model yielded promising results when applied to a set of variables stars observed by the mission \textit{Hipparcos}. One entire class, the RRAB, was set apart as our fictive ``new'' class. The other classes were divided into a training and a test set; the RRAB were mixed into this test set, to play the role of novelties. In the first step, we trained the TSDM on the training set (without RRAB). Although the Dirichlet mixtures in the TSDM achieved somewhat worse classification accuracy than the classic random forest reference model  \citep{breiman2001random, dubath2011random}, our classifier was still able to capture the physical reality of our variable star data: all of its mistakes originated in real similarities of physics driving the variability of the different types, or in existing phenomenological similarities in the observable features of the stars.  When using the TSDM model as the background for the next, semi-supervised step which defines the FB model, it enabled us to detect our ``new'' class with a 86.11\% accuracy and a 99.79\% specificity, which is very promising.

In future research, we plan to allow for additional flexibility by extending the new-class model to have a prior for the number of components.  Post-processing according to some suitable criterion may furthermore be used to assess whether model~(\ref{ncmodel}) just contains one class or whether there are several new classes which had not been considered in the original labelled data.  See \cite{malsiner-walli2017postprocessing} and references therein. Our proposal differs from \cite{browne2011model} in that our data are provided in two steps: first a labeled training data set and then an unlabelled test set to be classified.  Furthermore, the number of components of the inner mixture of \cite{browne2011model} is known, which is unknown in our case.

There is scope for further improvement of the TSDM model. First, in the second stage of the TSDM model for classification (Section~\ref{tsdmfitting}), we discussed the priors that can be used in the second stage. Though in our applications we used a non-informative prior, we can set a subjective prior that will represent the prior beliefs about the distribution of our parameters. As alluded to earlier, this will require a close collaboration with domain experts, but will improve our model and increase the flexibility.  Second, we used Dirichlet densities as a natural choice for modeling data in the probability scale. However, further investigation needs to be done to see how a two-stage copula mixture model would perform, as opposed to the TSDM model. Copulas hold an advantage over Dirichlet distributions that dependencies among random variables can be modeled as well \citep{embrechts2001modelling}. Finally, in the new-class model component of the FB model, we fixed the number of components in the beginning before fitting the new class to the data. In practice, it would be worthwhile not to fix it in the beginning to allow more flexibility. The next step could be to add a prior on the number of components so that the FB model will be able to choose the number of components from the data.

From a practical point of view, it would be interesting to explore the model in a Bayesian context. We can use concepts of conjugate exponential family. Mixture of Dirichlet distributions is an exponential family distribution, hence availing conjugate priors for our model, and subsequently the posterior distribution of the function of the parameter of interest. Results in \cite{diaconis1979conjugate} can be used to infer on the parameters of interest. Also to infer on the posterior, we can take advantage of the fact that the mixture of Dirichlet distributions is a member of the conjugate-exponential family since  (i)  the complete data likelihood is in the exponential family and (ii) the parameter prior is conjugate to the complete data likelihood. Variational Bayesian learning algorithms can be used to maximize the posterior of our model \citep{ghahramani2000variational}.

\color{black}

\section*{Acknowledgments}
This project was supported by SID 2018 grant ``Advanced statistical modelling for indexing celestial objects'' (BIRD185983) awarded by the Department of Statistical Sciences of the University of Padova.

\appendix

\section{Complete-data likelihoods}

\subsection{TSDM model}\label{app:TSDM}
Let us focus upon class $k$.  Let $J_k$ be the number of Dirichlet components of the corresponding mixture density (\ref{tsdmeqn-inner}).  Given the training set $\mathbf{y}=(\mathbf{y_1},\ldots,\mathbf{y_{n_k}})$ of dataset which belong to class $k$, denote by $\mathbf{s^T}=(s_1,\ldots,s_{n_k})$ the corresponding set of unobserved variables which indicate from which particular Dirichlet distribution, among the $J_k$ given ones, the observations $\mathbf{y_i}$ arise.  Each $s_i$, $i=1,\ldots,n_k$, can take the values $1,\ldots,J_k$ with probability $\pi_{k1},\ldots,\pi_{kJ_k}$.  

To apply the EM algorithm we need the complete-data likelihood \linebreak $L(\boldsymbol{\pi_k},\boldsymbol{\alpha_k};\mathbf{y},\mathbf{s})=f(\mathbf{y},\mathbf{s};\boldsymbol{\pi_k},\boldsymbol{\alpha_k})$, where $\boldsymbol{\alpha^T_k}=(\boldsymbol{\alpha_{k1}^T},\ldots,\boldsymbol{\alpha_{kJ_k}^T})$.  Had we observed the values of $\mathbf{s}$, the contribution from a single data point $(\mathbf{y_i},s_i)$ to the likelihood would be 
$$ f(\mathbf{y_i},s_i;\boldsymbol{\pi_k},\boldsymbol{\alpha_k}) =
\prod_{j=1}^{J_k}\left[\pi_{kj}f(\mathbf{y_i};\boldsymbol{\alpha_{kj}})\right]^{\mathbf{I}(s_i=j)}, $$
where $\mathbf {I}(s_i=j)$
%
%
is an indicator function as defined at (\ref{indicatorS}).  The complete-data log likelihood takes hence the form 
\begin{equation}\label{complete-data-lik}
\ell(\boldsymbol{\pi_k},\boldsymbol{\alpha_k};\mathbf{y},\mathbf{s}) = 
\sum_{i=1}^{n_k} \sum_{j=1}^{J_k}\mathbf{I}(s_i=j)\left[\log \pi_{kj} + \log f(\mathbf{y_i};\boldsymbol{\alpha_{kj}})\right]. 
\end{equation}
In the E-step of the EM algorithm we have to take the expectation of (\ref{complete-data-lik}) with respect to the conditional density $f(\mathbf{s}\mid \mathbf{y};\boldsymbol{\pi_k^\prime},\boldsymbol{\alpha_k^\prime})$ given that $\mathbf{Y} = \mathbf{y}$ for the parameter value $(\boldsymbol{\pi_k^\prime},\boldsymbol{\alpha_k^\prime})$.  Setting $\boldsymbol{\theta^T}=(\boldsymbol{\pi_k^T},\boldsymbol{\alpha_k^T})$, this yields the objective function
\begin{equation}\label{Q-function}
Q(\boldsymbol{\theta};\boldsymbol{\theta^\prime}) =   
\sum_{i=1}^{n_k} \sum_{j=1}^{J_k}w_{kj}(\mathbf{y_i};\boldsymbol{\theta^\prime})\left[\log \pi_{kj} + \log f(\mathbf{y_i};\boldsymbol{\alpha_{kj}})\right],\
\end{equation}
where 
$$
w_{kj}(\mathbf{y_i};\boldsymbol{\theta^\prime}) = 
\frac{\pi^\prime_{kj}f(\mathbf{y_i};\boldsymbol{\alpha_{kj}^\prime})}{\sum_{j=1}^{J_k}\pi^\prime_{kj}f(\mathbf{y_i};\boldsymbol{\alpha_{kj}^\prime})}
$$
are the conditional probabilities $\text{Pr}(S_i=j \mid \mathbf{Y_i} = \mathbf{y_i}; \boldsymbol{\theta_k^\prime})$ that the data point $\mathbf{y_i}$ arises from the $j$th Dirichlet distribution of class $k$.  The M-step of the EM algorithm maximizes $Q(\boldsymbol{\theta};\boldsymbol{\theta^\prime})$ over $\boldsymbol{\theta}$ for the fixed value $\boldsymbol{\theta^\prime}$.  This leads to the closed form solution
$ \pi_{kj} = n_k^{-1} \sum_{i=1}^{n_k} w_{kj}(\mathbf{y_i};\boldsymbol{\theta^\prime}) $
for the mixing probabilities $\pi_{kj}$.  The estimates of the additional parameters $\boldsymbol{\alpha_k}$ are obtained from the weighted log likelihood which forms the second term of (\ref{Q-function}).

\subsection{FB model}\label{app:fb}
Let $J_{K+1}$ be the number of Dirichlet components of the new-class mixture density (\ref{ncmodel}).  Let $\mathbf{y}=(\mathbf{y_1},\ldots,\mathbf{y_{n}})$ be the dataset that need be classified as either belonging to the background model $f_B(\cdot)$ or as arising from one of the $J_{K+1}$ Dirichlet distributions which form the new-class model $f_{NC}(\cdot)$.  Denote by $\mathbf{s^T}=(s_1,\ldots,s_{n})$ the corresponding set of unobserved variables which indicate from which part of the model the observations $\mathbf{y_i}$ stem.  Each $s_i$, $i=1,\ldots,n$, can take the values $0,1,\ldots,J_{K+1}$ with probabilities $\lambda_0,\lambda_1\ldots,\lambda_{K+1}$.  

To apply the EM algorithm we need the complete-data likelihood \linebreak $L(\boldsymbol{\lambda},\boldsymbol{\beta};\mathbf{y},\mathbf{s})=f(\mathbf{y},\mathbf{s};\boldsymbol{\lambda},\boldsymbol{\beta})$, where $\boldsymbol{\beta^T}=(\boldsymbol{\beta_{1}^T},\ldots,\boldsymbol{\beta_{J_{K+1}}^T})$.  Had we observed the values of $\mathbf{s}$, the contribution from a single data point $(\mathbf{y_i},s_i)$ to the likelihood would be 
$$ f(\mathbf{y_i},s_i;\boldsymbol{\lambda},\boldsymbol{\beta}) =
\lambda_0f_B(\mathbf{y_i})^{\mathbf{I}(s_i=0)} \times \prod_{j=1}^{J_{K+1}}\left[\lambda_jf(\mathbf{y_i};\boldsymbol{\beta_{j}})\right]^{\mathbf{I}(s_i=j)}, $$
where $\mathbf {I}(s_i=j)$
%
%
is an indicator function as defined at (\ref{indicatorS}).  The complete-data log likelihood takes hence the form 
\begin{eqnarray}
\nonumber
\ell(\boldsymbol{\lambda},\boldsymbol{\beta};\mathbf{y},\mathbf{s}) & = &
\sum_{i=1}^{n} \mathbf{I}(s_i=0)\left[\log \lambda_0 + \log f_B(\mathbf{y_i})\right] \\
\label{complete-data-lik-2}
& + & \sum_{i=1}^{n} \sum_{j=1}^{J_{K+1}}\mathbf{I}(s_i=j)\left[\log \lambda_j + \log f(\mathbf{y_i};\boldsymbol{\beta_{j}})\right] . 
\end{eqnarray}
In the E-step of the EM algorithm we have to take the expectation of (\ref{complete-data-lik-2}) with respect to the conditional density $f(\mathbf{s}\mid \mathbf{y};\boldsymbol{\lambda^\prime},\boldsymbol{\beta^\prime})$ given that $\mathbf{Y} = \mathbf{y}$ for the parameter value $(\boldsymbol{\lambda^\prime},\boldsymbol{\beta^\prime})$.  Setting $\boldsymbol{\psi^T}=(\boldsymbol{\lambda^T},\boldsymbol{\beta^T})$, this yields the objective function
\begin{eqnarray}
\nonumber
Q(\boldsymbol{\psi};\boldsymbol{\psi^\prime}) 
& = & \sum_{i=1}^{n} w_{0}(\mathbf{y_i};\boldsymbol{\psi^\prime}) \left[\log \lambda_0 + \log f_B(\mathbf{y_i})\right]
\\
\label{Q-function-2}
& + & \sum_{i=1}^{n} \sum_{j=1}^{J_{K+1}}w_{j}(\mathbf{y_i};\boldsymbol{\psi^\prime})\left[\log \lambda_j + \log f(\mathbf{y_i};\boldsymbol{\beta_{j}})\right],\
\end{eqnarray}
where 
$$
w_0(\mathbf{y_i};\boldsymbol{\psi^\prime}) =
\frac{\lambda_0^\prime f_B(\mathbf{y_i})}{\lambda_0^\prime f_B(\mathbf{y_i})+\sum_{j=1}^{J_{K+1}}{\lambda}_j^\prime f(\mathbf{y_i};\boldsymbol{\beta{^\prime}_j})} 
$$
and
$$ w_j(\mathbf{y_i};\boldsymbol{\psi^\prime}) =
\frac{\lambda_j^\prime f(\mathbf{y_i};\boldsymbol{\beta^\prime_j})}{\lambda_0^\prime f_B(\mathbf{y_i})+\sum_{j=1}^{J_{K+1}}{\lambda}_j^\prime f(\mathbf{y_i};\boldsymbol{\beta^\prime_j})} 
$$
for $j=1,\ldots,J_{K+1}$, are the conditional probabilities $\text{Pr}(S_i=j \mid \mathbf{Y_i} = \mathbf{y_i}; \boldsymbol{\psi^\prime})$ that the data point $\mathbf{y_i}$ arises from the fixed background or from the $j$th Dirichlet distribution of the new-class model, respectively.  The M-step of the EM algorithm maximizes $Q(\boldsymbol{\psi};\boldsymbol{\psi^\prime})$ over $\boldsymbol{\psi}$ for the fixed value $\boldsymbol{\psi^\prime}$.  This leads to the closed form solution
$\lambda_{j} = n^{-1} \sum_{i=1}^{n} w_{j}(\mathbf{y_i};\boldsymbol{\psi^\prime}) $
for the mixing probabilities $\lambda_k$.  The estimates of the additional parameters $\boldsymbol{\beta}$ are obtained from the weighted log likelihood which forms the second term of (\ref{Q-function}).

\begin{table}[p]
\centering
\caption{ The complete list of the 42 attributes derived for the 1,661 periodic variable stars which belong to the 23 types considered in this work as taken from \cite{dubath2011random}.  The names in boldface identify the attributes used to validate our margin-free classification and new class detection model with, in brackets, their relative importance.
}
{\footnotesize
\begin{tabular}{lp{0.7\linewidth}} 
\toprule
Attribute name                & Attribute description \\ 
\midrule
(16) {\bf p2pScatterOnDetrendedTS} & measure of the point-to-point scatter of the time series of brightness after removing a slow polynomial trend \\
p2pScatterOnFoldedTS          & measure of point-to-point scatter of the sequence of brightness after finding the period and phase-folding  \\
scatterOnResidualTS           & square root of variance of the residuals after modelling \\
(10) {\bf Raw\_weightedStdDev}     & weighted standard deviation of the brightness \\
(6) {\bf Raw\_weightedSkewness}   & weighted skewness of the brightness \\
Raw\_weightedKurtosis         & weighted kurtosis of the brightness \\
(3) {\bf Raw\_percentileRange10}  & 0.1--quantile minus the median of the raw brightness   \\
stetsonJ                      & measure of correlation between closely spaced brightness values \\
stetsonJweighted              & measure of correlation between closely spaced brightness values \\
stetsonK                      & measure of correlation between closely spaced brightness values \\
WstetsonJ                     & measure of correlation between closely spaced brightness values \\
WstetsonJweighted             & measure of correlation between closely spaced brightness values \\
(14) {\bf WstetsonK}               & measure of correlation between closely spaced brightness values \\
logPnonQso                    & measure of stochastic variability of components in the light curve \\
logPqso                       & measure of stochastic variability of components in the light curve  \\
(13) {\bf qsoVar}                  & measure of stochastic variability of components in the light curve \\
(11) {\bf nonQsoVar}               & measure of stochastic variability of components in the light curve \\
(1) {\bf LogPeriod}               & decadic logarithm of the period in days  \\
(9) {\bf LogAmplitude}            & decadic logarithm of the peak-to-peak amplitude \\
HarmNum                       & the highest significant order of harmonic terms in a least squares model fit  \\
(8) {\bf A11}                     & amplitude of the first harmonic term    \\
A12                           & amplitude of the second harmonic term      \\
(15) {\bf PH12}                    & relative phase of the second harmonic term \\
A13                           & amplitude of the third harmonic term  \\
PH13                          & relative phase of the third harmonic term \\
A14                           & amplitude of the fourth harmonic term \\
PH14                          & relative phase of the fourth harmonic term \\
A15                           & amplitude of the fifth harmonic term \\
PH15                          & relative phase of the fifth harmonic term \\
(12) {\bf logA11minusA}            & $\log_{10}(1 + \mid A11 - \sqrt{\sum A1j^2}\mid)$ \\
logA12\_11                    & $\log_{10}(1 + A12/A11)$ \\
logA13\_12                    & $\log_{10}(1 + A13/A12)$  \\
absGlat                       & absolute value of Galactic latitude \\
Glat                          & Galactic latitude \\
Glon                          & Galactic longitude \\
Parallax                      & the parallax of the object (in milliarcsec) equivalent to distance     \\
Absolute\_ag00                & estimate of the absolute brightness of the object \\
(7) {\bf BV\_color}               & difference between apparent brightness measured in the astronomical B and V filters\\
(5) {\bf VI\_color}               & difference between apparent brightness measured in the astronomical V and I filters \\
(2) {\bf JmK}                     & difference between apparent brightness measured in the infrared J and K filters \\
(4) {\bf JmH}                     & difference between apparent brightness measured in the astronomical J and H filters \\
HmK                           & difference between apparent brightness measured in the astronomical H and K filters \\ 
\bottomrule
\multicolumn{2}{l}{}                                                                    
\end{tabular}
}
\label{table_allattributes}
\end{table}

\section{Data pre-processing}\label{datapreprocessing}
This section describes how we reduced the number of attributes from the original 42 listed in Table~\ref{table_allattributes} to the 16 used in the validation process.  We furthermore describe how these were suitably transformed onto the simplex.

\subsection{Selection of attributes}\label{selectionofattributes}
We initially sorted the 42 original attributes by decreasing value of the \textit{mean decrease in accuracy} measure provided by a random forests classifier \citep{breiman2001random}.  We then selected the smallest number of attributes which still provided high accuracy by applying a variant of the forward selection strategy adopted in \cite{dubath2011random}.  The resulting list was finally integrated with some of the highly correlated attributes which had been discarded at first but are of high astronomical value.  The final set included the 16 attributes highlighted in boldface in Table~\ref{table_allattributes}; the number in brackets gives their importance in decreasing order.

\subsection{Transformation to the simplex}\label{transformation2simplex}
The transformation onto the simplex of the original measurements of the $D-1=16$ attributes was done in two steps.  At first, we logit-transformed the empirical distribution function 
$$\hat F_n^j(x)=n^{-1}\sum_{i=1}^n I_{(-\infty, x]}(x_{ij}), \quad j=1,\ldots,16,$$ 
for the $n$ observations $x_{ij}$, $i=1,\ldots,n$, of attribute $j$, smoothed it using natural cubic splines, and finally back transformed it to the original scale to yield the predictions $\tilde F_{ij}=\tilde F_n^j(x_{ij})$.  This way we are able to inter- and extrapolate the empirical probability for new data which may lie outside the range formed by the training set.  The smoothed empirical probabilities $(\tilde F_{i1}, \ldots, \tilde F_{i16})$ of the $D-1$ attributes for observation $i=1,\ldots,n$ are summed up and subtracted from $D-1$ to yield an additional fake attribute $F_{i17}= 16-\sum_{j=1}^{16}\tilde F_{ij}$ which is appended to the original data vector.  This whole vector is divided by $D-1=16$ to yield the final vector $\big(y_{i1},y_{i2},\ldots,y_{i16},y_{i17}\big)$ of data points which belong to the simplex of dimension $D=17$.

\bibliographystyle{apalike}
\bibliography{Paper2018.bib}

\vspace{2cm}
\begin{table}[h]
\begin{center}
\begin{tabular}{cc}
{\bf Address of the first and second authors} & {\bf Address of the third author} \\
University of Padova & EPFL \\
Department of Statistical Sciences & Institute of Mathematics\\
Via Cesare Battisti, 241 & Station 8 \\
35121 Padova (PD), Italy & 1015 Lausanne (VD), Switzerland \\[2ex]
ucanreach.princejohn@gmail.com & maria.sueveges@gmail.com \\
alessandra.brazzale@unipd.it & \\
\end{tabular}
\end{center}
\end{table}

\end{document}